\documentclass[12pt]{article}
\usepackage{amsmath}
\usepackage[dvips]{graphicx}
\input amssym

\begin{document}
	
    \def\prg#1{\medskip\noindent{\bf #1}}
	\def\inn{\,\rfloor\,}                  \def\bt{{\bar\tau}}
	\def\br{{\bar\rho}}
	\def\prg#1{\medskip\noindent{\bf #1}}  \def\ra{\rightarrow}
	\def\lra{\leftrightarrow}              \def\Ra{\Rightarrow}
	\def\nin{\noindent}                    \def\pd{\partial}
	\def\dis{\displaystyle}                \def\inn{\,\rfloor\,}
	\def\grl{{GR$_\Lambda$}}               \def\Lra{{\Leftrightarrow}}
	\def\cs{{\scriptstyle\rm CS}}          \def\ads3{{\rm AdS$_3$}}
	\def\Leff{\hbox{$\mit\L_{\hspace{.6pt}\rm eff}\,$}}
	\def\bull{\raise.25ex\hbox{\vrule height.8ex width.8ex}}
	\def\ric{{(Ric)}}                      \def\tric{{(\widetilde{Ric})}}
	\def\tmgl{\hbox{TMG$_\Lambda$}}
	\def\Lie{{\cal L}\hspace{-.7em}\raise.25ex\hbox{--}\hspace{.2em}}
	\def\sS{\hspace{2pt}S\hspace{-0.83em}\diagup}   \def\hd{{^\star}}
	\def\dis{\displaystyle}     \def\ul#1{\underline{#1}}
	\def\pgt{{\scriptstyle\rm PGT}}
	\def\ul{{\underline}}
	
	\def\hook{\hbox{\vrule height0pt width4pt depth0.3pt
			\vrule height7pt width0.3pt depth0.3pt
			\vrule height0pt width2pt depth0pt}\hspace{0.8pt}}
	\def\semidirect{\;{\rlap{$\supset$}\times}\;}
	\def\first{\rm (1ST)}       \def\second{\hspace{-1cm}\rm (2ND)}
	\def\bm#1{\hbox{{\boldmath $#1$}}}      \def\un#1{\underline{#1}}
	\def\nb#1{\marginpar{{\large\bf #1}}}
	\def\diag{{\rm diag}\,}
	
	\def\G{\Gamma}        \def\S{\Sigma}        \def\L{{\mit\Lambda}}
	\def\D{\Delta}        \def\Th{\Theta}
	\def\a{\alpha}        \def\b{\beta}         \def\g{\gamma}
	\def\d{\delta}        \def\m{\mu}           \def\n{\nu}
	\def\th{\theta}       \def\k{\kappa}        \def\l{\lambda}
	\def\vphi{\varphi}    \def\ve{\varepsilon}  \def\p{\pi}
	\def\r{\rho}          \def\Om{\Omega}       \def\om{\omega}
	\def\s{\sigma}        \def\t{\tau}          \def\eps{\epsilon}
	\def\nab{\nabla}      \def\btz{{\rm BTZ}}   \def\heps{\hat\eps}
	\def\bu{{\bar u}}     \def\bv{{\bar v}}     \def\bs{{\bar s}}
	\def\te{{\tilde e}}   \def\tk{{\tilde k}}
	\def\bare{{\bar e}}   \def\bark{{\bar k}}
	\def\barom{{\bar\omega}}  \def\barg{{\bar g}}
	\def\tphi{{\tilde\vphi}}  \def\tt{{\tilde t}}
	\def\te{{\tilde e}}
	
	\def\tG{{\tilde G}}   \def\cF{{\cal F}}    \def\bH{{\bar H}}
	\def\cL{{\cal L}}     \def\cM{{\cal M }}   \def\cE{{\cal E}}
	\def\cH{{\cal H}}     \def\hcH{\hat{\cH}}
	\def\cK{{\cal K}}     \def\hcK{\hat{\cK}}  \def\cT{{\cal T}}
	\def\cO{{\cal O}}     \def\hcO{\hat{\cal O}} \def\cV{{\cal V}}
	\def\tom{{\tilde\omega}}  \def\cE{{\cal E}}     \def\tR{{\tilde R}}
	\def\cR{{\cal R}}    \def\hR{{\hat R}{}}   \def\hL{{\hat\L}}
	\def\tb{{\tilde b}}  \def\tA{{\tilde A}}   \def\tv{{\tilde v}}
	\def\tT{{\tilde T}}  \def\tR{{\tilde R}}   \def\tcL{{\tilde\cL}}
	\def\hy{{\hat y}}    \def\tcO{{\tilde\cO}} \def\hom{{\hat\om}}
	\def\he{{\hat e}}
	\def\nn{\nonumber}                    \def\vsm{\vspace{-10pt}}
	\def\be{\begin{equation}}             \def\ee{\end{equation}}
	\def\ba#1{\begin{array}{#1}}          \def\ea{\end{array}}
	\def\bea{\begin{eqnarray} }           \def\eea{\end{eqnarray} }
	\def\beann{\begin{eqnarray*} }        \def\eeann{\end{eqnarray*} }
	\def\beal{\begin{eqalign}}            \def\eeal{\end{eqalign}}
	\def\lab#1{\label{eq:#1}}             \def\eq#1{(\ref{eq:#1})}
	\def\bsubeq{\begin{subequations}}     \def\esubeq{\end{subequations}}
	\def\bitem{\begin{itemize}}           \def\eitem{\end{itemize}}
	\renewcommand{\theequation}{\thesection.\arabic{equation}}
	
\title{Maximally symmetric black holes with torsion in Lovelock gravity}
\author{ B. Cvetkovi\'c and D. Simi\'c\footnote{
		Email addresses: {cbranislav@ipb.ac.rs, dsimic@ipb.ac.rs}}\\
	Institute of Physics, University of Belgrade \\
	Pregrevica 118, 11080 Belgrade, Serbia}

\date{}
\maketitle

\begin{abstract}
We analyze the existence of static black hole solutions with maximally symmetric horizons and nontrivial torsion in five-dimensional Lovelock gravity. For each symmetry class, we obtain a black hole solution characterized by two independent constants. We also determine the conserved charges associated with these solutions.
\end{abstract}

\section{Introduction}
Lovelock gravity \cite{1,2} is distinguished among higher-curvature gravitational theories by the fact that its field equations remain second order, despite the presence of higher powers of the curvature in the Lagrangian. More recently, Lovelock gravity has also attracted attention in the context of dimensional regularization of gravity \cite{x4}.

This property makes Lovelock gravity a well-controlled framework in which questions such as how higher-curvature corrections modify black hole solutions \cite{3,4,5,6}, how they affect black hole entropy and thermodynamics, and what their holographic implications are can be investigated systematically.

Five dimensions (5D) constitute the lowest dimension in which higher-curvature terms contribute nontrivially to the dynamics of Lovelock gravity. Consequently, 5D Lovelock gravity provides a natural setting for addressing the questions mentioned above.

Another interesting feature of 5D Lovelock gravity is its rich topological structure, which allows for a variety of distinct topologies of the event horizon. This aspect has been extensively studied in the literature \cite{7,8,9,10,11,d1,x2,x3}, primarily in the torsion-free sector of the theory.

In addition to these properties, Lovelock gravity admits vacuum solutions with nontrivial torsion \cite{12,13,14}. This feature distinguishes it from Riemann–Cartan gravity, which provides the minimal first-order extension of general relativity.

In this work, we investigate vacuum black hole solutions of 5D Lovelock gravity, focusing on the simplest case of maximally symmetric horizons. More general horizon geometries and topologies are possible and will be considered in future work.

The paper is organized as follows. In Section 2, we provide a concise overview of Lovelock gravity. In Section 3, we construct static black hole solutions with maximally symmetric horizons. Section 4 is devoted to the computation and discussion of the corresponding conserved charges. In Section 5, we analyze the near-horizon limit of the extremal black hole solutions.

\section{Lovelock gravity}
\setcounter{equation}{0}
We provide a concise overview of Lovelock gravity and the relevant mathematical concepts. For further details, see \cite{1,15,16,17}.

The dynamical variables of Lovelock gravity are the vielbein $e^i$, a 1-form, and the spin connection $\om^{ij}=-\om^{ji}$, also a 1-form. In a coordinate chart with coordinates $x^\m$, these 1-forms can be written as $e^i=e^i{_\m}dx^\m$, $\om^{ij}=\om^{ij}_\m dx^\m$.  
Metric tensor is constructed from vielbein and flat metric $\eta_{ij}$
\bea
&&g=\eta_{ij}e^i\otimes e^j=g_{\m\n}dx^\m\otimes dx^\n \, ,\nn\\
&& g_{\m\n}=\eta_{ij}e^i{_\m}e^j{_\n}\, ,
\qquad \eta_{ij}=(+,-,-,\dots,-)\, ,                   \nn
\eea
where $\otimes$ stands for tensor product and symmetry of $\eta_{ij}$ enforces symmetry of $g_{\mu\nu}$.
 
From the vielbein and spin connection we can construct torsion $T^i$ and curvature $R^{ij}$ 2-forms as follows
\bea
&&T^i=\nabla e^i\equiv de^i+\om^i_{\ j}\wedge e^j
=\frac{1}{2}T^i{}_{\m\n}dx^\m\wedge dx^\n\, ,         \nn\\
&&R^{ij}=d\om^{ij}+\om^{i}_{\ k}\wedge\om^{kj}
=\frac{1}{2}R^{ij}{}_{\m\n}dx^\m\wedge dx^\n\, , \nn
\eea
where $\nabla=dx^\m\nabla_\m$ is the exterior covariant derivative and $\wedge$ is exterior product of differential forms which we won't write explicitly anymore.

\subsection{Action and equations of motion}
The Lagrangian of Lovelock gravity in $D$ dimensions in the first order formulation is constructed as the linear combination of the dimensionally continued Euler densities $L_p$. Euler densities are defined as 
\be
L_p=\ve_{i_1 i_2 \dots i_D} R^{i_1i_2} \dots R^{i_{2p-1} i_{2p}}e^{i_{2p+1}} \dots e^{i_D}.\nn
\ee
In 5$D$ there are three Euler densities and the action of Lovelock gravity is given by
\bea
I=\ve_{ijkln}\int\left(\frac{\a_0}5e^{i}e^{j}e^{k}e^{l}e^{n}+\frac{\a_1}3R^{ij}e^{k}e^{l}e^{n}+\a_2 R^{ij}R^{kl}e^{n}\right)\,. \lab{2.1}
\eea

Variation of the action with respect to vielbein $e^i$
and spin connection $\om^{ij}$ gives the field equations
\be
\ve_{ijkln}\left(\a_0e^{j}e^{k}e^{l}e^{n}+\a_1R^{jk}e^le^n+\a_2R^{jk}R^{ln}\right)=0\,,\lab{2.2}
\ee
and
\be
\ve_{ijkln}\left(\a_1 e^ke^l+2\a_2R^{kl}\right)T^n=0\,.\lab{2.3}
\ee
\section{Maximally symmetric static solutions}
\setcounter{equation}{0}
The most general metric of the maximally symmetric and static solution in five dimensions is 
\be
ds^2=N^2dt^2-B^{-2} dr^2-r^2d\Omega^2_{k}\,,
\ee
where $d\Omega^2_{k}$ is the metric on a three-dimensional maximally symmetric Euclidean manifold of constant curvature $k$, with $(k=1,0,-1)$,  corresponding to spherical, planar and hyperbolic geometries respectively.

For the three cases considered here, the corresponding metrics $d\Omega^2_{k}$ are given by
\begin{equation}
	d\Omega^2_{1}=d\psi^2+\sin^2 \psi d\theta^2+\sin^2 \psi \sin^2 \theta d\vphi^2\,,
\end{equation}
\begin{equation}
	d\Omega^2_{0}=dx^2+dy^2+dz^2\,,
\end{equation}
and
\begin{equation}
	d\Omega^2_{-1}=d\psi^2+\sinh^2 \psi d\theta^2+\sinh^2 \psi \sin^2 \theta d\vphi^2\,.
\end{equation}
We shall analyze each of the geometries individually and find the  corresponding solutions. 

\subsection{Spherically symmetric solution}
For the spherically symmetric case, we choose the vielbeins in the simplest  diagonal form
\bea
&&e^0=Ndt\,,\qquad e^1=B^{-1}dr\,,\qquad e^2=rd\psi\,,\qquad e^3=r\sin\psi d\th\,,\nn\\
&&e^4=r\sin\psi\sin\th d\vphi\,.
\eea
The most general form of spin connection, which is compatible with the spherical symmetry, can be written as
\bea
&&\omega^{01}=A_0dt+A_1dr\,,\qquad  \omega^{02}=A_2d\psi\,,\nn\\
&&\omega^{03}=A_2\sin\psi d\theta\,,\qquad\omega^{04}=A_2\sin\psi \sin\theta d\vphi\,,\nn\\
&&\omega^{12}=A_3d\psi\,,\qquad \omega^{13}=A_3\sin\psi d\theta\,,\nn\\
&&\omega^{14}=A_3\sin\psi \sin\theta d\vphi\,,\qquad \omega^{23}=\cos\psi d\theta+A_4\sin\psi \sin\theta d\vphi\,,\nn\\
&&\omega^{24}=-A_4\sin\psi d\theta+\cos\psi \sin\theta d\vphi \,,\qquad \omega^{34}=A_4d\psi+\cos\theta d\vphi\,,
\eea
where $A_i$ are arbitrary functions of radial coordinate.

Solving the field equations with the help of software packages \cite{x1} gives
\bea
&&A_0=\frac{6\a_0}{\a_1}r+C_0\,,\qquad A_1=A_2=A_3=0\,,\qquad \  A_4=\sqrt{1-\frac{2 \a_0}{\a_1}r^2}\nn\\
&&N=B=\sqrt{\frac{C_1}{r^6}-\frac{2 C_0}{7}r-\frac{3 \a_0}{2\a_1}r^2}\,,\lab{3.4}
\eea
with the following constraint on coefficients
\be
\a_2=\frac{\a_1^2}{12\a_0}\,. \lab{3.5}
\ee
The sector is the same as the one in which spherically symmetric solution was found.

For the metric function to be real for large values of $r$ addition constraint must hold
\begin{equation}
	\frac{\alpha_0}{\alpha_1}<0.
\end{equation}
This solution is a  generalization of an already known BH solution. For $C_0=0$, this solution reduces to the solution previously obtained in Ref. \cite{3}.
\subsection{Planar solution}
For the planar geometry, we choose the vielbeins as follows
\bea
&&e^0=Ndt\,,\qquad e^1=B^{-1}dr\,,\qquad e^2=rdx\,,\qquad e^3=rdy\,,\nn\\
&&e^4=rdz\,.
\eea
The most general spin connection compatible with planar symmetry is
\bea
&&\omega^{01}=A_0dt\,,\qquad \omega^{02}=0\,,\nn\\
&&\omega^{03}=0\,,\qquad\omega^{04}=0\,,\nn\\
&&\omega^{12}=A_1dx \,,\qquad \omega^{13}=A_1dy\,,\nn\\
&&\omega^{14}=A_1dz \,,\qquad \omega^{23}=A_2dz\,,\nn\\
&&\omega^{24}=-A_2dy\,,\qquad \omega^{34}=A_2dx\,,
\eea
where $A_i$ are arbitrary functions of radial coordinate.

The resulting solution is
\bea
&&A_0=\frac{6\a_0}{\a_1}r+C_0\,,\qquad A_1=0\,,\qquad \  A_2=\sqrt{-\frac{2 \a_0}{\a_1}r^2}\nn\\
&&N=B=\sqrt{\frac{C_1}{r^6}-\frac{2 C_0}{7}r-\frac{3 \a_0}{2 \a_1}r^2}\,,\lab{3.4}
\eea
with the following constraint on coefficients
\be
\a_2=\frac{\a_1^2}{12\a_0}\,. \lab{3.5}
\ee
The reality of the spin connection again requires
\begin{equation}
	\frac{\alpha_0}{\alpha_1}<0.
\end{equation}

\subsection{Hyperbolically symmetric solution}
For the hyperbolic geometry, the vielbeins are chosen as
\bea
&&e^0=Ndt\,,\qquad e^1=B^{-1}dr\,,\qquad e^2=rd\psi\,,\qquad e^3=r\sinh\psi d\th\,,\nn\\
&&e^4=r\sinh\psi\sin\th d\vphi\,.
\eea
The most general spin connection compatible with hyperbolic symmetry is
\bea
&&\omega^{01}=A_0dt+A_1dr \,,\qquad \omega^{02}=A_2d\psi\,,\nn\\
&&\omega^{03}=A_2\sinh\psi d\theta\,,\qquad\omega^{04}=A_2\sinh\psi \sin\theta d\vphi\,,\nn\\
&&\omega^{12}=A_3d\psi \,,\qquad \omega^{13}=A_3\sinh\psi d\theta\,,\nn\\
&&\omega^{14}=A_3\sinh\psi \sin\theta d\vphi \,,\qquad \omega^{23}=\cosh\psi d\theta+A_4\sinh\psi \sin\theta d\vphi\,,\nn\\
&&\omega^{24}=-A_4\sinh\psi d\theta+\cosh\psi \sin\theta d\vphi\,,\qquad \omega^{34}=A_4d\psi+\cos\theta d\vphi\,,
\eea
where $A_i$ are arbitrary functions of radial coordinate.

The field equations yield the following solution
\bea
&&A_0=\frac{6\a_0}{\a_1}r+C_0\,,\qquad A_1=A_2=A_3=0\,,\qquad \  A_4=\sqrt{-1-\frac{2 \a_0}{\a_1}r^2}\nn\\
&&N=B=\sqrt{\frac{C_1}{r^6}-\frac{2 C_0}{7}r-\frac{3 \a_0}{2\a_1}r^2}\,,\lab{3.4}
\eea
with the same constraint on coefficients as before
\be
\a_2=\frac{\a_1^2}{12\a_0}\,. \lab{3.5}
\ee
For the metric function to be real for large values of $r$ additional constraint must hold
\begin{equation}
	\frac{\alpha_0}{\alpha_1}<0.
\end{equation}
An additional constraint arises from the requirement that the spin connection be real-valued outside the horizon.
\begin{equation}
	-1-\frac{2 \a_0}{\a_1}r_h^2\geq 0\,.
\end{equation}
The spin connection is automatically real for the other two solutions, since the coefficients $\a_0$ and $\a_1$ have opposite signs.

\subsection{Existence of a horizon}

The solution contains two arbitrary constants, but it is not immediately clear whether a horizon with $r_h>0$ exists for an arbitrary choice of these parameters. We therefore identify the parameter ranges for which a horizon exists.

There are two horizons with $r_h>0$ when the parameters satisfy
\bea
C_0>0\,, \qquad C_1>0\,,\qquad C_1<-\frac{\alpha_1^7 C_0^8}{28(6\alpha_0)^7}\,,
\eea
The two horizons merge into a single degenerate horizon when
\begin{equation}
	C_1=-\frac{\alpha_1^7 C_0^8}{28(6\alpha_0)^7}\,.
\end{equation}
A single horizon exists when
\be
C_1<0\,,
\ee
with $C_0$ otherwise unrestricted.

In the special case when 
\bea
C_1=0\,,\qquad C_0>0\,,
\eea
there is a single horizon located at
\be
r_h=-\frac{4 \a_1 C_0}{21\a_0}\,.
\ee
Other parameter combinations lead to solutions with no horizon, indicating the presence of naked singularities. To determine this, we need to examine the relevant curvature and torsional invariants.
\subsection{Invariants}
\prg{Riemannian Invariants.}
The Riemannian Ricci scalar
\begin{equation}
	\tilde{R}=-\frac{30\alpha_0}{\alpha_1} +\frac{12C_1}{r^8} - \frac{24 C_0}{7r}\,,
\end{equation}
while the Riemann Kretschmann scalar is
\begin{equation}
	\tilde{R}_{\mu\nu\rho\sigma}\tilde{R}^{\mu\nu\rho\sigma}=\frac{6}{49}\left(\frac{735\alpha_0^2}{\alpha_1^2}+ \frac{12 C_0^2}{r^2}+ \frac{84\alpha_0(-7C_1+2C_0 r^7)}{\alpha_1 r^8} +\frac{28C_1(581 C_1+4 C_0 r^7)}{r^{16}}\right)\,.
\end{equation}
Both invariants have the same form for all three solutions. They are regular everywhere except at $r=0$, where they diverge unless $C_0=C_1=0$.
Thus, in the absence of a horizon, the solutions represent naked singularities.

\prg{Cartan curvature invariants.}
The scalar Cartan curvature is constant and has the same value for all three solutions
\be
R=-\frac{24\a_0}{\a_1}\,,
\ee
while the quadratic Cartan curvature invariant has the same functional form for all three solutions
\be
R^{ij}\wedge\hd R_{ij}=\frac{12\a_0^2}{7\a_1^2(2\a_0r^2-\a_1)}\left(133\a_0r^2+4\a_1C_0r-56\a_1-\frac{14\a_1C_1}{r^6}\right)e^0e^1e^2e^3e^4\,.
\ee
This quadratic invariant is regular everywhere except at $r=0$, where it diverges unless $C_1=0$.

\prg{Torsional invariants.}
For all three solutions we obtain
\be
T^i\wedge\hd T_i=6\left(\frac {7\a_0}{\a_1}+\frac{4C_0}{7r}-\frac2{r^2}-\frac{2C_1}{r^8}\right)e^0e^1e^2e^3e^4\,.
\ee
The quadratic torsion invariant is regular everywhere except at $r=0$, where it diverges unless $C_1=0$.

These invariants show that only the Riemannian invariants distinguish the solutions with $C_1=0$ as black holes when a horizon is present, or as naked singularities when no horizon exists.

\section{Conserved charges of the black holes}
\setcounter{equation}{0}

There are several methods for calculating the conserved charges associated with a Killing vector. In our case, the simplest approach is to apply Nester’s formula \cite{21}. Before proceeding with the calculation of the charges, we first introduce the necessary terminology.

The covariant momenta derived from the Lovelock action \eq{2.1} are given by
\bea
&&\t_i:=\frac{\pd L}{\pd T^i}=0\,,\\
&&\r_{ij}=\frac{\pd L}{\pd R^{ij}}=2\ve_{ijkln}\left(\frac {\a_1}{3}e^ke^l+2\a_2 R^{kl}\right)e^n\,.
\eea
We need to specify a reference spacetime with respect to which the conserved charges are measured. The appropriate choice is the spacetime with zero horizon radius, $r_h=0$, which corresponds to $C_0=C_1=0$. We also introduce the difference between any variable $X$ and its reference value $\bar X$, denoted by $\D X=X-\bar X$. 
The conserved charges $Q_\xi$ associated with a Killing vector $\xi$ are given by the quasi-local surface integral
$$
Q_\xi=\int_{\pd\S}B\,,
$$
where the boundary $\pd\S$ is located at infinity. For fields with suitable asymptotic behavior, the corresponding boundary term is given by \cite{21}
\be\lab{4.1}
B=(\xi\inn e^i)\D \t_i + \D e^i(\xi\inn\bt_i)
+\frac 12(\xi\inn\om^i{_j})\D\r_i{^j}
+\frac 12\D\om^i{_j}(\xi\inn\br_i{^j})\, ,
\ee
where $\inn$ denotes contraction.

From the expressions for the spin connection, it is straightforward to see that all components except $\omega^{01}$ have the same form as in the background. For $r_h\neq 0$, we have  $\omega^{ij}=\bar{\omega}^{ij}+C_0 dt (\delta^i_0\delta^j_1-\delta^j_0\delta^i_1)$, which means that the only non-vanishing difference  is $\D\om^{01}=C_0dt$.

Therefore, using that $\t_i=0$, formula \eq{4.1} takes the following simpler form
\be
B=\frac 12(\xi\inn\om^{ij})\D\r_{ij}+\frac 12\D\om^{01}(\xi\inn\br_{01})\,.
\ee
However, $\r_{01}$ vanishes due to the relation among the coefficients defining the sector in which the solutions exist. Therefore, the boundary term further simplifies to
\be
B=\frac 12(\xi\inn\om^{ij})\D\r_{ij}\,.
\ee
Using the Killing vectors given in Appendix A, we obtain, after straightforward calculations
\be
Q_{\xi_0}=Q_{\xi_1}=Q_{\xi_2}=Q_{\xi_3}=Q_{\xi_4}=Q_{\xi_5}=Q_{\xi_6}=0\,.
\ee
Thus, all conserved charges vanish for all three black hole solutions derived in this work; see Appendix B for details of the calculations. Owing to the static nature of the solutions, it is expected that all charges other than the energy vanish. The vanishing energy, or mass, of the black holes is a rather unusual property, although such solutions are known in the literature \cite{3,22}. Massless black hole solutions also occur in the closely related four-dimensional Chamseddine gravity \cite{23}.

\section{Near-horizon geometry}
\setcounter{equation}{0}

A near-horizon geometry can be associated with an extremal black hole for which the temperature vanishes, $T=0$. In the present case, extremality is equivalent to the metric function $N^2$ having a double zero at the horizon $r=r_h$.

The extremality conditions obtained from the metric function $N^2$ are
\bea
C_1=-\frac{\alpha_1^7 C_0^8}{28(6\alpha_0)^7}\,,\qquad  r_h=-\frac{\alpha_1 C_0}{6\alpha_0},
\eea
which fixes one of the two arbitrary constants, $C_1$, in terms of $C_0$.

The near-horizon geometry is obtained by introducing the coordinate transformation
\bea
r=r_h+\epsilon \rho,\ & t=\frac{\tau}{\epsilon},
\eea
while keeping the remaining coordinates unchanged, and subsequently taking the limit
\be
\epsilon\rightarrow 0.
\ee
This procedure is designed to yield a well-defined near-horizon limit of the metric, but it can fail in the presence of torsion \cite{25}. Since the solutions derived in this paper have nontrivial torsion, the existence of a regular near-horizon limit is not guaranteed. In particular, the regularity of the spin connection must also be checked \cite{c1}. Whenever the limit is regular, it describes a solution of the same theory as the original black hole solution. We analyze the three cases separately, emphasizing their common features and highlighting their differences.

\subsection{Spherically symmetric}
For the spherically symmetric black hole, the metric function in the near-horizon limit takes the simple form
\be
N(r)\rightarrow N(\rho)=\frac{\rho}{\ell}\,,
\ee
where
\be
\ell^2=-\frac{\alpha_1}{6\alpha_0}\,.
\ee
The near-horizon metric is
\be
ds^2=\frac{\rho^2}{\ell^2}d\tau^2-\frac{\ell^2}{\rho^2}d\rho^2-r_h^2(d\psi^2+\sin^2 \psi d\theta^2+\sin^2 \psi \sin^2 \theta d\vphi^2)\,,\lab{3.1}
\ee
The nonzero functions entering the spin connection have the following near-horizon limits:
\bea
A_0(r)\rightarrow A_0(\rho)=\frac{6\alpha_0\rho}{\alpha_1}\,,\qquad  A_4(r)\rightarrow A_4(\rho)=\sqrt{1-\frac{2\alpha_0}{\alpha_1}r_h^2}\,,
\eea 
which are regular. This demonstrates that the original black hole geometry admits a regular near-horizon limit.

The nonzero components of the spin connection in the near-horizon limit are
\bea
&&\omega^{01}=-\frac{\rho}{\ell^2}d\tau\,,\qquad \omega^{24}=-\sqrt{1+\frac{\ell^2C_0^2}{3}}\sin\psi d\theta+\cos\psi \sin\theta d\vphi\,,  \nn\\
&&\omega^{23}=\sqrt{1+\frac{\ell^2C_0^2}{3}}\sin\psi \sin\theta d\vphi\,,\qquad \omega^{34}=\sqrt{1+\frac{\ell^2C_0^2}{3}}d\psi+\cos\theta d\vphi\,.
\eea
The metric shows that the near-horizon geometry is the direct product of two-dimensional anti-de Sitter space, $AdS_2$, and a three-dimensional sphere, $S^3$, of fixed radius $r_h$. The spin connection shows that torsion is present in both the $AdS_2$ and $S^3$ sectors.

\subsection{Planar}

For the planar case, the metric function is the same as in the spherically symmetric case, while the near-horizon limit of the spin connection is different. The near-horizon metric is
\begin{equation}
	ds^2=\frac{\rho^2}{\ell^2}d\tau^2-\frac{\ell^2}{\rho^2}d\rho^2-r_h^2(dx^2+dy^2+dz^2)\,.
\end{equation}
The corresponding vielbein can be read off directly from this metric.

The nonzero components of the near-horizon spin connection are
\bea
\omega^{01}=-\frac{\rho}{\ell^2}d\tau\,, & \omega^{23}=\frac{r_h}{\ell\sqrt{3}}dz\,,\nn\\
\omega^{24}=-\frac{r_h}{\ell\sqrt{3}}dy\,,& \omega^{34}=\frac{r_h}{\ell\sqrt{3}}dx\,.
\eea
Thus, the near-horizon geometry is the direct product
$AdS_2\times R^3$, where the three-dimensional Euclidean space has a constant metric scale set by $r_h$. Both factors carry nontrivial torsion.

\subsection{Hyperbolic}

For the hyperbolic case, the metric function is again the same, and the near-horizon metric is
\be
ds^2=\frac{\rho^2}{\ell^2}d\tau^2-\frac{\ell^2}{\rho^2}d\rho^2-r_h^2(d\psi^2+\sinh^2 \psi d\theta^2+\sinh^2 \psi \sin^2 \theta d\vphi^2)\,,\lab{3.1}
\ee
The most important difference arises from the near-horizon limit of the function $A_4$ appearing in the spin connection
\begin{equation}
	A_4\rightarrow\sqrt{-1-\frac{\alpha_1C_0^2}{18\alpha_0}}=\sqrt{-1+\frac{\ell^2C_0^2}{3}}\,.
\end{equation}
For the spin connection to remain real-valued, this requires
\begin{equation}
	\ell^2C_0^2\geq3\,,
\end{equation}
Otherwise, the spin connection is not a real-valued differential form.

The nonzero components of the spin connection in the near-horizon geometry are
\bea
&&\omega^{01}=-\frac{\rho}{\ell^2}d\tau\,, \qquad \omega^{24}=-\sqrt{-1+\frac{\ell^2C_0^2}{3}}\sinh\psi d\theta+\cosh\psi \sin\theta d\vphi\,,  \nn\\
 &&\omega^{23}=\sqrt{-1+\frac{\ell^2C_0^2}{3}}\sinh\psi \sin\theta d\vphi\,,\qquad \omega^{34}=\sqrt{-1+\frac{\ell^2C_0^2}{3}}d\psi+\cos\theta d\vphi\,.
\eea
Geometrically, the near-horizon geometry is the direct product $AdS_2\times H^3$,
where $H^3$ is a three-dimensional hyperbolic space of fixed radius $r_h$.
\section{Concluding Remarks}
\setcounter{equation}{0}
We derived static black hole solutions with maximally symmetric horizons and nontrivial torsion in five-dimensional Lovelock gravity. More specifically, we analyzed three possible geometries corresponding to spherical, planar, and hyperbolic spatial sections. The solutions are characterized by two independent integration constants.

A distinguishing feature of these solutions is the presence of torsion, which is not an additional structure imposed on a Riemannian black hole geometry, but an essential part of the solutions themselves. The spin connection contains nontrivial torsional contributions in all three symmetry sectors, and their form depends on the topology of the black hole horizon.

An important property of the derived black hole solutions is that all of them have vanishing energy as well as vanishing values of all other conserved charges. Massless black hole solutions are already known in five-dimensional Lovelock gravity, and their existence is related to the special structure of Lovelock dynamics in five dimensions. In the present case, our results show that the existence of a black hole geometry with nonvanishing torsion does not necessarily imply a nonzero mass.

The action evaluated on the black hole solutions vanishes as a consequence of the relation among the coefficients in the Lagrangian that determines the sector in which the solutions exist. This provides an alternative way of seeing that both the conserved charges and the entropy vanish.

We also investigated the extremal limit of the black hole solutions, which is possible because the solutions are characterized by two independent constants. The extremality condition fixes one of these constants in terms of the other. After taking the near-horizon limit, we find that the resulting geometries possess a universal two-dimensional anti-de Sitter factor. More precisely, the near-horizon geometries take the forms
\bea
AdS_2\times S^3,\qquad
AdS_2\times R^3,\qquad
AdS_2\times H^3,
\eea
for spherical, planar, and hyperbolic horizons, respectively. The near-horizon limit preserves the torsional nature of the original black hole solutions.

These results demonstrate that five-dimensional Lovelock gravity admits a class of black hole geometries in which torsion plays an essential role. The fact that the same metric function corresponds to different horizon topologies, while the spin connection distinguishes between them, provides an interesting setting for studying the interplay between torsion and horizon geometry in theories with torsion.

The stability of these solutions and the behavior of perturbations around the black hole backgrounds deserve further investigation. The near-horizon geometries derived here may also provide a starting point for the classification of extremal solutions with torsion in five-dimensional Lovelock gravity. Finally, it would be interesting to investigate whether these black hole solutions can be generalized to rotating or boosted configurations.

\section{Acknowledgments}
This work was partially supported by the Ministry of Science, Technological development and
Innovations of the Republic of Serbia.

\appendix

\section{Killing Vectors}
\setcounter{equation}{0}
Besides timelike Killling vector $\xi_0=\partial_t$ that all static solutions share each solution has 6 additional Killing vectors.
\subsection{Spherically symmetric}
Killing vector that correspond to SO(4) symmetry are as follows:
\bea
&&\xi_{1}=\cos\th\pd_\psi-\cot\psi\sin\th\pd_\th\,,\nn\\
&&\xi_{2}=\sin\th\cos\vphi\pd_\psi+\cot\psi\cos\th\cos\vphi\pd_\th-\frac{\cot\psi}{\sin\th}\sin\vphi\pd_\vphi\,,\nn\\
&&\xi_{3}=\sin\th\sin\vphi\pd_\psi+\cot\psi\cos\th\sin\vphi\pd_\th+\frac{\cot\psi}{\sin\th}\cos\vphi\pd_\vphi\,,\nn\\
&&\xi_{4}=\cos\vphi\pd_\th-\cot\th\sin\vphi\pd_\vphi\,,\nn\\
&&\xi_{5}=\sin\vphi\pd_\th+\cot\th\cos\vphi\pd_\vphi\,,\nn\\
&&\xi_{6}=\pd_\vphi\,.
\eea

\subsection{Hyperbolically symmetric}
List of Killing vector of SO(3,1) symmetry is as follows:
\bea
&&\xi_{1}=\cos\th\pd_\psi-\coth\psi\sin\th\pd_\th\,,\nn\\
&&\xi_{2}=\sin\th\cos\vphi\pd_\psi+\coth\psi\cos\th\cos\vphi\pd_\th-\frac{\coth\psi}{\sin\th}\sin\vphi\pd_\vphi\,,\nn\\
&&\xi_{3}=\sin\th\sin\vphi\pd_\psi+\coth\psi\cos\th\sin\vphi\pd_\th+\frac{\coth\psi}{\sin\th}\cos\vphi\pd_\vphi\,,\nn\\
&&\xi_{4}=\cos\vphi\pd_\th-\cot\th\sin\vphi\pd_\vphi\,,\nn\\
&&\xi_{5}=\sin\vphi\pd_\th+\cot\th\cos\vphi\pd_\vphi\,,\nn\\
&&\xi_{6}=\pd_\vphi\,.
\eea
Killing vectors $\xi_1$, $\xi_2$ and $\xi_3$ correspond to boosts while $\xi_4$, $\xi_5$ and $\xi_6$ stem from rotations.

\subsection{Planar(Euclidean symmetry)}
Euclidean symmetry $E(3)$ is semi-direct product of translation group $T^3$ and rotations $SO(3)$. Killing vectors are given by
\bea
&&\xi_{1}=\pd_x\,,\nn\\
&&\xi_{2}=\pd_y\,,\nn\\
&&\xi_{3}=\pd_z\,,\nn\\
&&\xi_{4}=y\pd_x-x\pd_y\,,\nn\\
&&\xi_{5}=z\pd_x-x\pd_z\,,\nn\\
&&\xi_{6}=z\pd_y-y\pd_z\,.
\eea
Killing vectors $\xi_1$, $\xi_2$ and $\xi_3$ are associated to translations while $\xi_4$, $\xi_5$ and $\xi_6$ correspond to rotations.

\section{Calculation of charges}
\setcounter{equation}{0}
\subsection{Spherically and hyperbolically symmetric}
We treat spherically and hyperbolically symmetric solution simultaneously because the only important difference is change of trigonometric into hyperbolic function. We use $\cot(h)\psi$ to label $\cot\psi$ in spherical and $\coth\psi$ in hyperbolic case respectively.

The form of the covariant momenta for both spherical and hyperbolic solution is the same and given by
\bea
&&\r_{01}=\frac{4\left(\a_1^2-12\a_0\a_2\right)}{\a_1}e^2e^3e^4\equiv 0\,,\qquad \r_{02}=-\frac{8\a_1}3 e^1e^3e^4\,,\qquad \r_{03}=\frac{8\a_1}3 e^1e^2e^4\,,\nn\\
&&\r_{04}=-\frac{8\a_1}3 e^1e^2e^3\,,\qquad  \r_{12}=\frac{8\a_1}3e^0e^3e^4-\frac{4\a_1N}{3A_4}e^0e^1e^2\,,\nn\\
&&\r_{13}=-\frac{8\a_1}3 e^0e^2e^4-\frac{4\a_1N}{3A_4}e^0e^1e^3\,,\qquad\r_{14}=\frac{8\a_1}3 e^0e^2e^3-\frac{4\a_1N}{3A_4}e^0e^1e^4\,,\nn\\
&&\r_{23}=0\,,\qquad \r_{24}=0\,,\qquad \r_{34}=0\,.
\eea

The explicit expressions for charges are as follows: 

\bea
&&Q_{\xi_0}=\int_{\pd \S}\om^{01}{}_t\D \r_{01}=0\,,\nn\\
&&Q_{\xi_1}=\int_{\pd \S}-\cot(h)\psi\sin\th\left(\om^{23}{}_\th\D\r_{23}+\om^{24}{}_\th\D\r_{24}\right)=0\,,\nn\\
&&Q_{\xi_2}=\int_{\pd\S}\cot(h)\psi\cos\th\cos\vphi\left(\om^{23}{}_\th\D\r_{23}+\om^{24}{}_\th\D\r_{24}\right)\nn\\
&&-\frac{\cot(h)\psi}{\sin\th}\sin\vphi\left(\om^{23}{}_\vphi\D\r_{23}+\om^{24}{}_\vphi\D\r_{24}+\om^{34}{}_\vphi\D\r_{34}\right)=0\,,\nn\\
&&Q_{\xi_3}=\int_{\pd\S}\cot(h)\psi\cos\th\sin\vphi\left(\om^{23}{}_\th\D\r_{23}+\om^{24}{}_\th\D\r_{24}\right)\nn\\
&&+\frac{\cot(h)\psi}{\sin\th}\cos\vphi\left(\om^{23}{}_\vphi\D\r_{23}+\om^{24}{}_\vphi\D\r_{24}+\om^{34}{}_\vphi\D\r_{34}\right)=0\,,\nn\\
&&Q_{\xi_4}=\int_{\pd\S}\cos\vphi\left(\om^{23}{}_\th\D\r_{23}+\om^{24}{}_\th\D\r_{24}\right)\nn\\
&&-\cot\th\sin\vphi\left(\om^{23}{}_\vphi\D\r_{23}+\om^{24}{}_\vphi\D\r_{24}+\om^{34}{}_\vphi\D\r_{34}\right)=0\,,\nn\\
&&Q_{\xi_5}=\int_{\pd\S}\sin\vphi\left(\om^{23}{}_\th\D\r_{23}+\om^{24}{}_\th\D\r_{24}\right)\nn\\
&&+\cot\th\cos\vphi\left(\om^{23}{}_\vphi\D\r_{23}+\om^{24}{}_\vphi\D\r_{24}+\om^{34}{}_\vphi\D\r_{34}\right)=0\,,\nn\\
&&Q_{\xi_6}=\int_{\pd\S}\left(\om^{23}{}_\vphi\D\r_{23}+\om^{24}{}_\vphi\D\r_{24}+\om^{34}{}_\vphi\D\r_{34}\right)=0\,.
\eea
In the calculation first we determine all non-vanishing contractions of Killing vector with spin connection. That leads to the expressions written above. Then we note that all $\D \r$ appearing are zero because those components of $\r$ vanish, consequently all charges are zero.
\subsection{Planar}
The covariant momenta for planar solution are given by
\bea
&&\r_{01}=\frac{4\left(\a_1^2-12\a_0\a_2\right)}{\a_1}e^2e^3e^4\equiv 0\,,\qquad \r_{02}=-\frac{8\a_1}3 e^1e^3e^4\,,\qquad \r_{03}=\frac{8\a_1}3 e^1e^2e^4\,,\nn\\
&&\r_{04}=-\frac{8\a_1}3 e^1e^2e^3\,,\qquad  \r_{12}=\frac{8\a_1}3e^0e^3e^4+\frac{2\a_1^2N}{3\a_0r^2A_2}e^0e^1e^2\,,\nn\\
&&\r_{13}=-\frac{8\a_1}3 e^0e^2e^4+\frac{2\a_1^2N}{3\a_0r^2A_2}e^0e^1e^3\,,\qquad\r_{14}=\frac{8\a_1}3 e^0e^2e^3+\frac{2\a_1^2N}{3\a_0r^2A_2}e^0e^1e^4\,,\nn\\
&&\r_{23}=0\,,\qquad \r_{24}=0\,,\qquad \r_{34}=0\,.
\eea
Notice that covariant momenta are almost of the same form as for spherical and hyperbolic case.
The charges are evaluated as follows:
\bea
&&Q_{\xi_0}=\int_{\pd \S}\om^{01}{}_t\D \r_{01}=0\,,\nn\\
&&Q_{\xi_1}=\int_{\pd \S}\om^{34}{}_x\D\r_{34}=0\,,\nn\\
&&Q_{\xi_2}=\int_{\pd\S}\om^{24}{}_y\D\r_{24}=0\,,\nn\\
&&Q_{\xi_3}=\int_{\pd\S}\om^{23}{}_z\D\r_{23}=0\,,\nn\\
&&Q_{\xi_4}=\int_{\pd\S}\left( y\om^{34}{}_x\D\r_{34}-x\om^{24}{}_y\D\r_{24}\right)=0\,,\nn\\
&&Q_{\xi_5}=\int_{\pd\S}\left( z\om^{34}{}_x\D\r_{34}-x\om^{23}{}_z\D\r_{23}\right)=0\,,\nn\\
&&Q_{\xi_6}=\int_{\pd\S}\left(z\om^{24}{}_y\D\r_{24}-y\om^{23}{}_z\D\r_{23}\right)=0\,.
\eea
Again, we first calculated all non-zero contractions as given above. The charges are zero because all $\D\r$ multyplying non-zero $\omega$ components are zero, for the same reasons as in other two cases.

\end{document}